# Towards Eco-friendly Database Management Systems

Willis Lang and Jignesh M. Patel
Computer Sciences Department
University of Wisconsin-Madison, USA
{wlang, jignesh}@cs.wisc.edu

## ABSTRACT

Database management systems (DBMSs) have largely ignored the task of managing the energy consumed during query processing. Both economical and environmental factors now require that DBMSs pay close attention to energy consumption. In this paper we approach this issue by considering energy consumption as a first-class performance goal for query processing in a DBMS. We present two concrete techniques that can be used by a DBMS to directly manage the energy consumption. Both techniques trade energy consumption for performance. The first technique, called PVC, leverages the ability of modern processors to execute at lower processor voltage and frequency. The second technique, called QED, uses query aggregation to leverage common components of queries in a workload. Using experiments run on a commercial DBMS and MySQL, we show that PVC can reduce the processor energy consumption by 49% of the original consumption while increasing the response time by only 3%. On MySQL, PVC can reduce energy consumption by 20% with a response time penalty of only 6%. For simple selection queries with no predicate overlap, we show that QED can be used to gracefully trade response time for energy, reducing energy consumption by 54% for a 43% increase in average response time. In this paper we also highlight some research issues in the emerging area of energy-efficient data processing.

## 1. INTRODUCTION

Servers consume enormous amounts of energy. A recent report [11] showed that when considering only the servers in data centers, in 2005 an estimated 1.2% of the total U.S. energy consumption is attributed to powering and cooling servers, at an estimated cost of $2.7B. Another report by the EPA [1] estimates that in 2006, the servers and data centers in the US alone consumed about 61 billion kilowatt-hours (kWh) at a cost of $4.5 billion, which is about 1.5% of the total U.S. electricity consumption. If the current methods for powering servers and data centers continue, then it is predicted that this energy consumption will nearly double by 2011. Other recent work [10] highlight that energy cost is the third largest cost in a data center (after server hardware, and power distribution and cooling costs). Energy management is quickly becoming an important metric and design criteria for modern data center management and planning [2, 10]. More broadly, energy efficiency is an emerging critical design and operational criteria for computing environments that includes data centers, small clusters, and even stand-alone servers. DBMSs running in server environments have largely ignored energy efficiency, but we can no longer afford such oversight. Both economical *and* environmental factors require that we start considering energy as a critical "performance" metric in DBMSs.

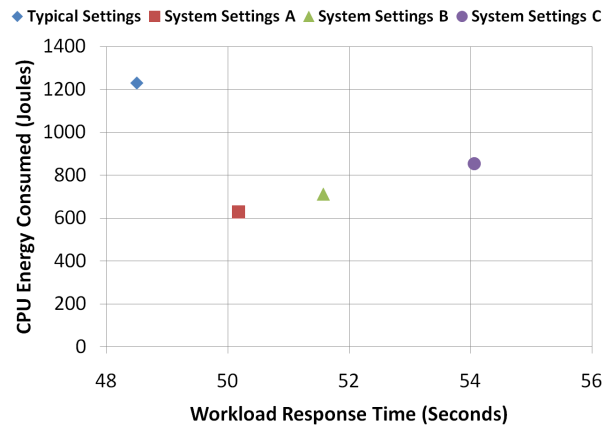

**Figure 1: TPC-H Query 5 on a Commercial DBMS**

This paper describes a new project, called ecoDB, in which we are developing energy efficient data processing techniques. The key focus of this project is on examining techniques in which energy can be traded for performance. While four decades of research in the database and distributed data processing communities has produced a wealth of methods that optimize for response time and/or throughput, the addition of energy consumption considerations opens a wide range of research issues. We will explore these new issues in the ecoDB project. As a starting point, we consider using "energy consumption" as a first-class metric in a DBMS when planning and processing queries.

We note that the overall goal of ecoDB is to investigate energy efficient methods for general data processing in distributed computing environments (which includes servers, clusters, and data centers, running either DBMSs or some other data processing system, such as MapReduce [5]). There are two broad classes of techniques that can be used to improve the overall energy efficiency in such data processing environments. "Global" techniques can be used to change some aspect of how the entire system is managed or used. An example of such a technique is to change the job scheduling method for the entire system to achieve better energy efficiency. The second class of techniques is "local" techniques that can be used to improve the energy efficiency of methods which are used for processing data at individual nodes. The focus of this paper is on the "local" techniques, and in an DBMS environment.

We begin our investigation with a simple model for trading energy and performance in a DBMS. Consider a system that can determine, on a per query or workload basis, power settings given response time and energy consumption goals. Figure 1 shows a sample plot of an actual commercial DBMS running a workload consisting of TPC-H queries. This figure is plotted from *real* data that we have collected under settings that we discuss in detail in Section 3.3.

Figure 1 shows a number of alternative ways to evaluate a database workload. The top left point in this figure corresponds to a traditional operating point, which runs the workload in about 48.5 seconds. At this point the CPU energy consumption is just over 1200 Joules. An alternative operating point for this workload is to use the system setting A, which has a response time penalty of about 3%, but improves the CPU energy efficiency by 49% (over a conventional/typical setting). A system that does not have strict response time requirements, could choose to run this workload using system setting A to save energy. The other points, setting B and C, are worse than system setting A as they result in longer response times and consume more energy. (See Section 3.3 for a more detailed discussion.)

Two interesting questions quickly follow from this discussion: "*How does a system generate graphs as shown in Figure 1?*" and "*How can such a graph be used?*" We only offer partial answers to these questions in this paper. As the reader will see each partial answer raises new interesting questions, each of which is worthy of focused follow-on work, pointing to the far reaching promise of energy efficiency as a new area for database research.

These two questions lay the foundation for performance versus energy efficiency tradeoffs. For a DBMS to generate Figure 1, it must be aware of system hardware capabilities and operating characteristics, and take that into account during query optimization and evaluation to produce this plot. The DBMS will need to accurately estimate and continuously measure the hardware energy consumption characteristics under both static and dynamic loads. To add to the challenge, many hardware components are implementing hardware-based energy efficiency methods, and the DBMS and the hardware must co-operate to manage energy consumption. It is essential that hardware expose appropriate controls to allow the DBMS to provide directives related to energy consumption (otherwise issues similar in spirit to those raised in [16] can also happen for energy management).

Once the DBMS is able to gather such data and compose a representative operating plot, it must have some systematic method to shift between different settings. The DBMS must be able to make automatic transitions given protocols provided by administrators and hardware load sensors. Factors such as Service Level Agreements (SLAs) may restrict the choices. A data center operating near peak may have no choice but to aim for the fastest query response time. However, when the data center is not operating at peak capacity (which is the common case [7]) it may have the option of using an operating point that can save energy. (An interesting related issue is how to take into account workload power characteristics, such as Figure 1, and to work backward to create viable parameters for an SLA.) It may also be interesting to consider cases where our initial prediction for energy consumption are incorrect and then to dynamically adapt our query plan midflight to meet our response time and energy goals.

The focus of this paper is on mechanisms that can be used to create graphs as in Figure 1, which can trade energy for performance. While there are a number of mechanisms here (again pointing to the broad research promise of this area), we only consider two mechanisms in this paper.

The first mechanism, called **P**rocessor **V**oltage/Frequency **C**ontrol (**PVC**), lets the DBMS explicitly change the processor voltage and frequency parameters. Modern processors have been driven to consider energy efficiency as a first-class design goal very aggressively over the last few years (which as a side-effect has resulted in the multi-core processor era). Processors can now take commands from software to adjust their voltage/frequencies to operate at points other than their peak performance. This control is in addition to automatic internal transitions made by the processor that move the processor to lower performance states when deemed necessary (such as when the processor is idle, or underutilized, or current system settings require a lower performance level).

The second mechanism that we consider is at the workload management level. Consider a DBMS in which there is some "admission control" for running queries. A query is examined, perhaps even optimized, before it is allowed to run in the system. Now consider maintaining a queue of queries that are waiting to run. If the queries in the queue have some common components, such as common subexpressions, it might make sense from the pure performance perspective to use multi-query optimization techniques to optimize the workload. Now consider exploiting this generic mechanism to explicitly delay queries for workloads in which there are often common components across different queries (and delays can be tolerated). The explicit delay allows queries to build up in the queue, which then allows multi-query optimization methods to evaluate the entire batch more efficiently. Consequently the average per-query energy consumption can be lowered. We call this technique "Improved **Q**uery **E**nergy-efficiency by Introducing Explicit **D**elays" (**QED**).

We have implemented PVC and QED and tested them on an actual system with both a leading free database engine, MySQL (which was chosen because of its popularity), and a commercial DBMS. Our results are promising and show that PVC can be used to reduce the CPU energy consumption by 20% and 49%, while incurring 6% and 3% response time penalties on MySQL and the commercial DBMS respectively. On a workload with simple selection queries, on MySQL QED saves 54% of the CPU energy consumption while increasing the average query response time by 43%.

We believe that this is the first paper that proposes concrete methods that can be used by DBMSs to move towards eco-friendly query processing. Our framework for trading energy for performance opens a new area of research, and our early experimental results in this area demonstrate the viability of this line of thinking.

The remainder of this paper is organized as follows: In the next section, we discuss some opportunities for energy efficiency in query processing environments. Sections 3 and 4 describe and evaluate the PVC and QED methods respectively. Related work is covered in Section 5, and finally Section 6 contains our concluding remarks.

## 2. OPPORTUNITIES FOR ENERGY EFFICIENCY

In a database server environment there are a number of points in the overall landscape that present opportunities for improved energy efficiency. For example, in a data center, the power supply and power distribution are very inefficient, and can often lead to losses as high as 44% [10]. Techniques exist today to improve this efficiency and are often affordable (and given the intense focus on this inefficiency, manufacturers are aggressively designing newer, more efficient techniques). Other points also exist in data centers to improve energy efficiency. Data centers typically operate at low loads most of the time [2, 10] and moving to higher utilization can save energy (though issues such as data replication make this challenging). Changing and managing the cooling technology can also lead to dramatic savings in energy consumption [9, 10, 13]. Choosing the hardware carefully, by building more balanced systems and using slower but more energy-efficient hardware, is another way of reducing the energy consumption of data centers [10].

These improvements in "best-practices" will eventually happen and improve the energy efficiency of data centers. Once that happens, an interesting next point for improved energy efficiency for processing DBMS workloads is the way in which the DBMS explicitly makes choices to improve energy efficiency (note at that point the DBMS component of the total energy consumption is likely to be larger!). Traditional database investigations into improving query response times will still be valuable, as faster query times often means lower energy consumption. However, the move towards more energy efficient data centers, coupled with the fact that most data centers are underutilized most of the time, opens up opportunities for methods that trade energy for performance. These tradeoffs can be investigated at various levels including at the operator-level (e.g. rethinking join algorithms in this context), query-level (e.g. considering the effect of different query plans for the energy versus response time tradeoff), workload management per server (e.g. considering alternatives to scheduling), and to workload management for the entire collection of servers (e.g. scheduling and using techniques to turn entire servers off when not required). We believe each of these topics lead to interesting directions for future work in this area. However, many of these techniques will require hardware that is more amenable to quick changes in performance states.

If one looks at the hardware components inside a server, most components are not very energy efficient. It has been bemoaned that modern hardware consumes more than half the peak energy even when idle [2]. However, a large reason for this behavior is because energy has not been a first-class design goal for most hardware components. But, this is changing. Take two big sources of energy consumption on server motherboards: processor and memory. Processors have gotten significantly more energy efficient over the past five years (while adding more processing power). The processor manufacturers have done this by operating at lower voltages and also putting in smart mechanisms in the processor that automatically push the processor to lower performance states when it is idle. Memory has also slowly started getting more energy efficient over the last few years. Many high-density DRAM manufacturers now tout the energy efficiency of their products (many of these products also achieve energy efficiency by operating at lower voltages). Furthermore, there are efforts to build hardware that will put memory banks into deep sleep (thereby saving energy) when not in use.

Essentially, one can expect that future hardware will likely be more energy efficient than the hardware available today. Future hardware components, like the processors today, will likely also detect when they are idle and power down. However, these techniques will need to be complimented by techniques that provide more direct manipulation of the hardware components to manage energy efficiently, primarily because the higher-level software has better information about the job characteristics. As stated in [15], "responsibility of saving power lies with both system and user level software". Thus we expect techniques, such as those described in this paper, will likely be even more relevant in the future as energy efficiencies at various levels in server environments evolve towards removing the existing large wasteful drains on energy in current computing environments.

## 3. EXPLOITING PROCESSOR ENERGY EFFICIENCY TECHNIQUES

Energy has become a crucial aspect in the design of modern processors. Processors use a technique called dynamic voltage and frequency scaling (DVFS) to automatically reduce the processor voltage/frequency, moving the processor to lower power/performance states (p-states). Intel's SpeedStep and AMD's PowerNow! and Cool'n'Quiet are examples of built-in automatic processor DVFS. However, many software packages allow programs explicit access to different p-states. P-states are characterized by the combination of CPU multiplier and CPU voltage settings. The CPU frequency is a product of the front side bus (FSB) speed and the CPU multiplier, where the CPU multiplier is dictated by the p-state. The CPU voltage is based on the CPU multiplier, and a lower multiplier allows the CPU to operate at a lower voltage.

In addition to p-states, users have the ability to more finely tune the CPU and the entire system speed by changing the FSB speed. Overclocking enthusiasts largely use this technique to make the system go faster. But, this technique can destabilize the system, and can reduces the life span of the processor and other hardware components. However, the converse, which is the ability to underclock the system by slowing the FSB speed, does not reduce component lifespan (it may do the reverse and increase the component's lifespan).

It is important to distinguish the difference between CPU frequency modulation through p-state transitioning and underclocking. Traditional methods of CPU power management through p-state manipulation put a hard upper limit on the top p-state that a CPU can achieve.

For example, consider a CPU on a 333MHz FSB, with p-state multipliers of 9, 8, 7, and 6 and some corresponding decreasing voltages. When p-state control is used to manage power, the max p-state can be capped to a value. Lets assume a capped value of 7. This means that the top frequency the CPU can now achieve is 2.3Ghz ($= 7 \times 333$MHz), instead of 3Ghz ($= 9 \times 333$MHz).

In this paper, we use underclocking because instead of capping the multiplier, the FSB speed is now decreased which allows a finer granularity of CPU frequency modulation. Instead of dropping the frequency **by** 333Mhz with each multiplier cap, we simply modulate the base factor (the FSB speed) that the multiplier acts on.

With underclocking, we also retain the full capabilities of p-state transitions (which allows the CPU to reduce its energy consumption dynamically). In our example, capping the multiplier at 7 would mean that only 2 transitioning states are left available with an FSB speed of 333Mhz. Underclocking retains *all* the multiplier settings while globally reducing the frequency of all available p-states. This means we have all 4 settings corresponding to multiplier values ranging from 6 through 9, but these values are multiplied against a smaller (slower) FSB speed.

Consequently, lowering the FSB speed is a finer grained tool for changing the CPU frequency. Underclocking also affects other components connected to the Northbridge hub. Main memory is on the Northbridge, and its operating frequency is a multiple of the FSB (usually a different multiplier from the CPU multiplier).

Thus, underclocking also slows the main memory, which in turn reduces the amount of energy consumed by main memory. We examine the effects of these smaller frequency modulations through underclocking on the energy consumption of the CPU.

| PSU | MOBO | CPU | 1G RAM | 2G RAM | GPU | SYS ON | Measured |
|---|---|---|---|---|---|---|---|
| ✓ | ✓ | x | x | x | x | x | 9.2W |
| ✓ | ✓ | x | x | x | x | ✓ | 20.1W |
| ✓ | ✓ | ✓ | x | x | x | ✓ | 49.7W |
| ✓ | ✓ | ✓ | ✓ | x | x | ✓ | 54.0W |
| ✓ | ✓ | ✓ | ✓ | ✓ | x | ✓ | 55.7W |
| ✓ | ✓ | ✓ | ✓ | ✓ | ✓ | ✓ | 69.3W |

**Table 1: System Power Breakdown.** The motherboard is denoted as MOBO, CPU includes fan, SYS ON means that we have turned the system on by pressing the power button on the case.

### 3.1 System Under Test

The system that we use in this paper has the following main components: ASUS P5Q3 Deluxe Wifi-AP motherboard, Intel Core2-Duo E8500, $2\times 1G$ Kingston DDR3 main memory, ASUS GeForce 8400GS 256M, and a Western Digital Caviar SE16 320G SATA disk. The power supply unit (PSU) used was a Corsair VX450W PSU, which is labeled as an energy-efficient PSU under 80plus.org. System power draw was measured by a Yokogawa WT210 unit (as suggested by SPEC power benchmarks). The operating system used was Microsoft Windows Server 2008. All client applications accessing the database were written in Java 1.6 using JDBC connection drivers to the different database systems we used.

To measure the CPU power consumption, we used a hardware sensor provided by the motherboard manufacturer. The ASUS motherboard has an EPU processor that directly measures the CPU power. The ASUS P5Q3 Deluxe 6-Engine software displays information gathered from this onboard hardware sensor. Unfortunately, the only display mechanism is a GUI that shows the current CPU wattage. As a workaround, we recorded the CPU wattage by graphically sampling the GUI every second throughout the execution of the workload. CPU joules was recorded as the average sampled wattage multiplied by the workload execution time. There are drawbacks to using this method as the refresh rate of the 6-Engine is about 1 second. Furthermore, since we allowed Intel Speedstep to act freely, the CPU power fluctuates during an actual run. To address these issues, we create workloads that contain 10 queries for a given type of query. Each of these 10 queries has input predicate ranges that don't overlap. Then we take measurements for the entire workload, which is usually many minutes long. Finally, we run each workload five times and discard the top and bottom readings, and average the middle three readings. This average is reported in our results.

In all our experiments, we did not create any database indices.

### 3.2 CPU Energy Consumption

In this paper, we focus on the energy consumption of the CPU for the following two reasons: a) CPU employs sophisticated energy management techniques, and still draws a significant amount of power (see below), and b) other components are "primitive" in their energy characteristics and it is expected that these components will become more energy efficient over time. Including these other components would be interesting, but getting detailed and accurate hardware measurements on these other components on an actual motherboard is non-trivial. We have measured the energy characteristics of the hard disk drive and we report these in Section 3.5.

Since software services and operating systems that are always running make power breakdowns like the one we present next difficult, the results below are obtained without a hard disk and operating system to provide the most accurate per component power measurements. We measured the power drawn from the wall as we built up our machine, starting with just the PSU and the motherboard. These measurements are reported in Table 1.

Simply powering a motherboard with the PSU draws 9.24W *with the system off*. The PSU and motherboard themselves draw 20.13W when the board is turned on. We expect motherboards to increase in energy efficiency as some manufacturers such as ASUS and Gigabyte have already started to release boards with such efficiency as a marketed feature. (One of the reasons we chose the P5Q3 Deluxe was because it is touted as a leading "green" board.)

When the CPU is installed on the motherboard with the stock CPU fan, the power draw more than doubles. However, it is likely that when the CPU is installed, other components on the motherboard are activated, thus drawing more power.

The DDR3 main memory draws about 6W for 2 DIMMs (1GB each). In addition, while we show the GPU cost here, typical database servers may not need this component.

Finally, we estimate that the power efficiency of the PSU is around 83%, given the near 20% load exhibited of our system [6]. Thus, the results in Table 1 contain a signifiant amount of PSU losses, since we measured the power consumed by the entire system.

In the future, as other components besides the CPU are optimized for power, we can expect the CPU to continue to contribute towards a significant portion of the total energy cost. (Getting accurate measurements for other hardware components when the system is running and broadening this study is part of our future work.) We have also observed that the actual CPU power consumption when running the experiments (so the disk is connected and the OS and DBMS is running) is often about 25% of the overall system power consumption.

### 3.3 PVC

The **P**rocessor **V**oltage/Frequency **C**ontrol (**PVC**) technique uses the underclocking technique described above to trade energy for performance. We now present an experimental evaluation of PVC. For our tests, we use TPC-H as our workload. To keep the experiments manageable, we only ran TPC-H Q5. This query has a response time that is often close to the geometric mean of the power tests in many published results. This query has a six table join and a group by clause on one attribute [17]. We ran TPC-H on both a commercial DBMS and MySQL 5.1.28.

To explore the energy savings resulting from underclocking the system, we first run the workload in "stock setting" – which corresponds to no underclocking and represents the traditional way of processing queries. Then, we used the 6-engine software provided by ASUS to underclock the FSB by 5%, 10%, and 15%. In addition, we used this software to downgrade the CPU voltage into its preset "small" and "medium" voltage downgrades. We also ran the ASUS PC Probe II program, which continuously monitors the CPU settings and warns when the system settings result in instability. We found that using the "small" and "medium" CPU voltage downgrades, the system operated without any warnings.

Other motherboard settings were as follows: CPU loadline was set to "light," chipset voltage downgrade was set to "on," and CPU fan settings was set to "bios setting." Thus, for each workload, we have 7 CPU power results: stock (no altered settings) and the 3 underclocking settings, for the 2 CPU voltage downgrades. Unless stated otherwise, results will be presented as a ratio compared to the "stock" CPU and motherboard settings.

The results for the commercial DBMS are shown on the first page in Figure 1. These results are for the TPC-H workload with scale factor of 1.0. Each workload consists of ten Q5 queries with predicates using regions 'Asia' and 'America' and all five possible date ranges. Given the uniform nature of TPC-H, all ten queries in the workload perform the same amount of work, and have nonoverlapping predicates.

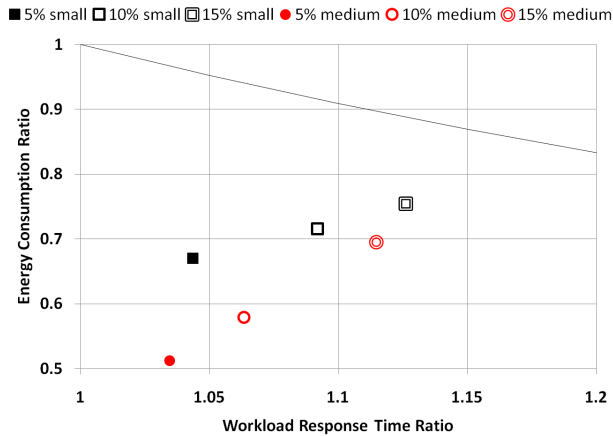

Figure 2: TPC-H Query 5 on a Commerical DBMS

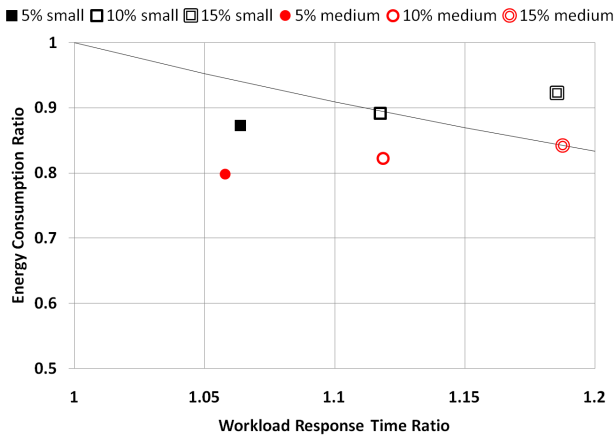

Figure 3: TPC-H Query 5 on MySQL

In Figure 1 the settings A, B, and C correspond to 5%, 10%, and 15% underclocking with "medium" voltage downgrade, and typical setting corresponds to the "stock" reading. As can be seen from Figure 1 the 5% setting dramatically reduces the energy (by 49%) for only a small drop in performance (3%). Underclocking beyond 5% actually increases the energy consumption. We delay the discussion of this issue to Section 3.4.

As noted earlier in Section 3.2, we are only measuring the CPU energy consumption. The effect on the overall system energy consumption is lower due to the poor power characteristics of existing hardware, which is likely to improve in the future. For example, in Figure 1 for the 5% underclocking with medium downgrade the overall system energy consumption only drops by 6%.

We also evaluated the effect of underclocking with "small" voltage setting. This result is shown in Figure 2. In this graph, the X-axis is the ratio of the CPU energy consumption with the modified settings divided by the CPU energy consumption in the stock setting. The Y-axis in this graph shows the ratio of the response time with the modified settings divided by the response time in the stock setting. Note that the stock setting is the origin, which is the top left point in this graph.

When comparing both energy and time, a useful metric to use is the Energy Delay Product (EDP), which is the product of the energy and the delay (this metric is discussed further in Section 3.4). The solid curve in Figure 2 corresponds to values where the EDP value is not changed. In other words, the change in energy consumed is matched by an equal change in the response time (in the opposite direction), and EDP remains constant. Lower EDP values are desirable in our settings as they crudely represent gaining a larger % of energy saving over the loss in response time. Consequently, points below the curve are "interesting".

From Figure 2 we see that the 5% underclocked system lowers the EDP for both the small and medium voltage downgrade settings, and the EDP with the medium voltage settings is lower than with the small voltage settings. (In Figure 2 the EDP for a data point is the shortest distance from the data point to the EDP curve.)

Figure 3 shows the plot for MySQL on TPC-H workload (scale 0.125). As before we plot the energy and response times as a ratio over the stock setting, and the origin at the top left corner corresponds to the stock setting. In this case, we used the memory storage engine of MySQL to stress the CPU.

Interesting in both Figures 2 and 3, underclocking beyond 5% actually worsens the EDP!

For example, in Figure 2, with "small" voltage downgrade, the EDP rises from -30% to -22% to -15% for 5%, 10%, and 15% underclocking respectively. For the "medium" voltage setting, in Figure 2, the EDP rises from -47% to -38% to -23% for the 5%, 10%, and 15% underclocking settings repectively.

In Figure 3, with the "small" voltage downgrade, EDP rises from -7% at 5%, then up goes to -0.4% and +9% at 10% and 15% respectively. Similarly, in Figure 3, for the "medium" voltage, EDP rises from -16% to -8% to 0% for 5%, 10%, and 15% underclocking respectively.

To understand this behavior, we need to dig deeper into the theoretical relationship between a processor's power consumption and its performance, which we discuss next.

### 3.4 Theoretical Modeling

It is a well known that circuit power can be modeled as $CV^2F$. $C$ is a constant factor, $V$ is the voltage, and $F$ is the frequency. We measured the voltage and frequency for the MySQL workload by continuous monitoring these values, and then averaging the observed values. Both voltage and frequency remained nearly constant for this workload (not surprising since the memory engine makes MySQL CPU-bound, rarely allowing the processor to go to lower p-states). The observed frequency was equal to $multiplier \times bus\_speed$.

We can calculate the EDP as $joules \times time$. Thus, $EDP = power \times time^2$. Since time is inversly proportional to frequency, we can reduce this equation to $EDP = CV^2/F$. Now, since $C$ is constant, the theoretical EDP is proportional to $V^2/F$.

Now let us compare our observed EDP from Figure 3 to the theoretical EDP described above. Figure 4 shows the observed EDP on the primary Y-axis, and the theoretical EDP on the secondary Y-axis (i.e. we simply plot $V^2/F$).

In the figure, we see that the observed EDP closely matches the theoretical model. Now, we can also understand why our observed energy costs increase when we underclock the system beyond 5%, while holding the CPU voltage downgrade steady — the additional execution time penalty simply overwhelms any CPU power gains due to additional underclocking.

### 3.5 Disk Access and Energy Consumption

So far our discussions have focused on trading CPU performance for energy consumption savings. In this section, we present data showing the energy consumption of the hard disk drive.

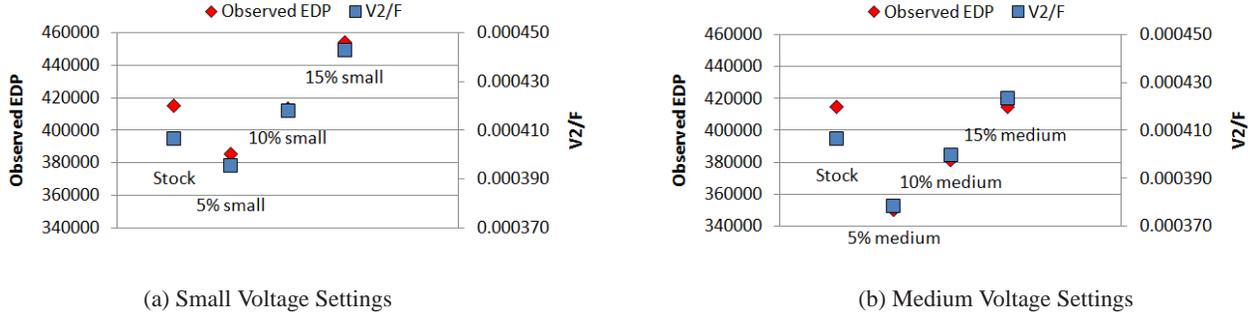

(a) Small Voltage Settings

(b) Medium Voltage Settings

Figure 4: A Comparison of the Observed EDP to the Theoretical $EDP = V^2/F$ Model

The hard disk drive in our SUT has two power lines – a 5V line and a 12 V line. We measured the current on both lines when running the TPCH Q5 workload using the commercial DBMS. We then computed the energy consumed on each line and summed up the energy consumption to compute the overall energy consumption of the hard disk drive.

The results shown in Figure 2 are for a warm database. Observation of the system during these runs found that the hard disk drive had significant activity even though the database was warm (and the size of the raw tables is less than the main memory capacity). When running the workload in stock settings, the CPU consumed 1228.7 Joules for the workload. During that time, the hard disk drive consumed 214.7 Joules of energy. This shows that the hard disk drive consumes about 1/6 of the energy of the CPU while running the workload on a warm database. We were then interested in observing what this ratio would look like on a cold system. We thus ran the same workload immediately following a system reboot. On a cold system, the workload took about three times longer to complete (156 seconds). Further, our CPU energy consumption was 2146.0 Joules while our hard disk now took 1135.4 Joules, namely more than half the energy used by the CPU. This result is not surprising as the disk now must fetch all necessary data from disk while the CPU may remain idle for extended periods of time.

We explored the disk energy consumption further by looking at differences in energy consumption between random and sequential access. For this analysis we took a 4GB file and read 1.6GB (400,000 4KB pages) of data from it using sequential and random accesses. For sequential access we simply read the first 1.6GB of the file. Random accesses were generated by computing random file pointer locations and seeking to that spot. Furthermore, we varied the amount/block of data that was fetched in each read call. We used the following block sizes: 4KB, 8KB, 16KB, and 32KB. In each run, we kept the amount of data that was read the same (so the number of calls made to read is half with 8KB page size compared to 4KB page size).

Figure 5 (a) shows the data throughput for these two accesses, for different read block sizes. In the figure, not surprisingly, we see that the sequential access throughput is constant regardless of the read size. For random access we see that throughput steadily rises as we increase the read block size, as by increasing the read block size we reduce the total number of disk seeks. In Figure 5 (b) we see the energy consumed per KB of data retrieved. Again, as expected, the energy cost for the sequential access is flat. Sequential access is more energy efficient per KB than random access, primarily because it is faster!.

Note that in Figure 5 the throughput and energy consumption changes at a slower rate than the rate of increase in the read block size. Increasing the read block size from 4KB to 8KB increased throughput and energy efficiency by about 1.88 times. Further increase in the read block size to 16KB and 32KB produces approximately 3.5 and 6 times improvements in both throughput and energy efficiency respectively. This shows that speed and energy efficiency performance is close but does not exactly follow the changes in the read block size.

These results suggest that the best way to save power in disk is to maintain as much sequential access as possible, i.e. reducing the IO (time) cost results in improved disk energy efficiency.

## 4. EXPLICIT QUERY DELAYS (QED)

In this section, we describe the QED method that can be used to reduce energy costs for a query workload. In this paper, we present this technique in a very simplistic setting.

Consider a simple workload model in which the database is presented with a series of single table select-only queries that are structurally the same but differ in the range of the selection predicate. Queries are issued to the database continuously one after the other with no delay between a query finishing and the next query arriving (i.e. think time is zero). Furthermore, we assume that the DBMS can only run one query in the system at a time. In a traditional DBMS this workload will result in each query being evaluated individually, and one after the other.

Now consider a different model of evaluating this workload that can trade average query response time for reduced overall energy consumption. In QED, queries are delayed and placed into a queue on arrival. When the queue reaches a certain threshold all the queries in the queue are examined to determine if they can be aggregated into a small number of groups, such that queries in each group can be evaluated together. For example, the select queries in our workload can be merged to a single group with a disjunction of the predicates in each query. This single query can then be run in the DBMS at a lower energy cost than the individual queries. QED also has a little bit of extra work to do with respect to splitting the result, which for this paper we do in the application logic and include the time and energy cost to do this extra computation.

To test QED, we created a simple workload with a series of single table queries with each query having a 2% selectivity based on the *l_quantity* attribute of the *lineitem* table of the TPC-H benchmark at 0.5 scaling factor. For this test, we use MySQL with its memory engine, and we run the system at "stock" system settings. Each 2% selectivity query has a predicate on a single *l_quantity* value drawn from the uniformly distributed 50 integer values. We run a simple workload where each query contains a different *l_quantity* predicate

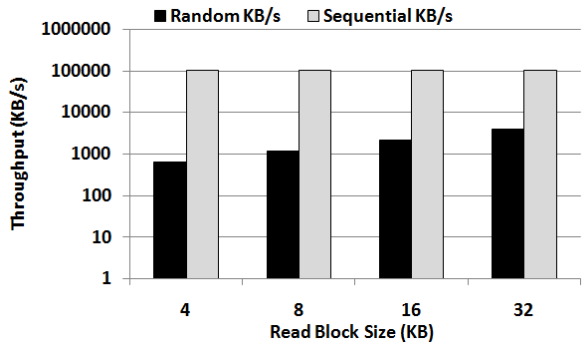
(a) Data Throughput for Random and Sequential Access.

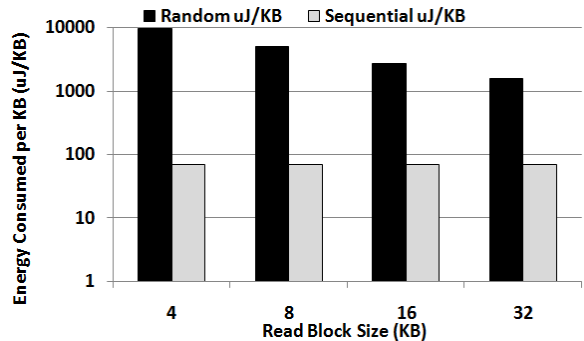
(b) Energy Consumption per KB for Random and Sequential Access

Figure 5: Hard Disk Energy Consumption for Different Data Access Patterns.

value (so there is no overlap amongst the selection predicates up to a batch size of 50). QED batches the queries and periodically sends an aggregated query to the system.

Figure 6 shows the per query response time and energy trade-off between running a set number of queries sequentially, or aggregating them together by predicate disjunction and issuing one large query (followed by running code that splits up the result to the individual queries). Here, we ran four different aggregation batch sizes, namely: 35, 40, 45, and 50 queries.

The time to run a batch of queries is measured from the time the batch of queries is issued to the database to the time the last query is returned. For the sequential scheme, this means that time and energy costs start as soon as the first query is sent to the DBMS. For QED, we do not count the time that it takes for the database to collect a batch of queries. Time and energy costs start when the batch of queries is sent to the DBMS. Essentially, we assume that the queue of queries builds up in a master system that is always on (perhaps there is a master for a group of servers) and that the DBMS machine goes to sleep when there is no work. (We admit that these assumptions need to be relaxed for a more realistic setting.)

QED improves energy efficiency at the cost of degrading the response time of the queries. Consider the point for batch size of 35. From Figure 6, we see that a batch size of 35 lowers the energy consumption by 46% while increasing the average response time by 52%. The EDP in this case is 18% lower compared to the traditional sequential approach. Increasing the batch size to 40 drops the energy consumption by 51% while response time increases by 50%. The EDP drop for the batch size of 40 is lowered by 26% (over the sequential case).

Going from a batch size of 35 to 50, as we keep increasing the batch size by 5, we see that the rate of change of the response time is constant, but the rate of change in the energy consumption is gradual, i.e. there is a diminishing decrease in energy consumption. A possible reason for this is that as we increase the number of queries, the amortized energy cost per sequential query decreases and so the relative benefit of QED per increase in batch size starts to diminish. In Figure 6, the largest batch size (of 50) results in the largest energy savings with the least amount of average response time degradation, which translates to the best EDP change.

We note that the response time degradation is most severe for the first query in the batch, and least for the last query in the batch. Furthermore, the degradation in response time for the first query increases as the batch size increases. A simple analytical model can be used to capture these effects in more detail, and can be used to consider the impact on SLAs.

We note that QED simply uses multi-query optimization [14] to evaluate a batch of queries more efficiently (than evaluating each individually). Essentially QED is exploiting the batch efficiency effects of multi-query optimization to improve response time per query, and consequently the energy consumption per query (though as we saw above the relationship is not strictly linear). Consequently generalization of our method to more complex workloads (beyond simple select queries) is feasible.

## 5. RELATED WORK

The need for energy efficiency and the high costs of energy consumption in data centers and server environments has been highlighted in various recent reports [1–4, 7, 11]. Recent work has also shown that the energy consumption of current hardware is not proportional to the actual load on the machine [2, 7], and current hardware consumes a significant amount of energy even when it is idle or near idle.

In the database community, JouleSort [12] is a recent addition to the sort benchmark that focuses on energy efficient sorting. This work has proposed a sort benchmark with three scales (for different sizes of input) and uses the number of sorted records per Joule as the benchmark metric. This work also examined various hardware systems, including laptops and servers, and showed that a custom machine with server capacity storage but with laptop disks and a desktop processor produces that best benchmark results.

Graefe's [8] paper presents high-level goals towards thinking about energy efficiency in database systems. As the paper says that it "represents a challenge rather than a solution" and presents various high-level ideas towards considering energy efficiency in DBMSs, no explicit proposal for trading energy for performance is contained. We wholeheartedly agree with the direction of that paper and welcome more proponents that push the community to consider energy efficiency in DBMSs.

To the best of our knowledge, our paper presents the first formulation of concrete methods to control the energy consumption in DBMSs and to consider techniques that trade energy for performance.

## 6. CONCLUSIONS AND DIRECTIONS FOR FUTURE WORK

In this paper, we have presented some proposal for considering energy efficient query processing in a database management system. Our proposals center around techniques that can trade energy for performance. We have also described two techniques that work within this framework. Actual experiments on two DBMSs demonstrate that these techniques achieve significant savings in energy, and often reduce the energy consumption by a larger amount compared to the degradation in response time (i.e. operate at lower EDPs).

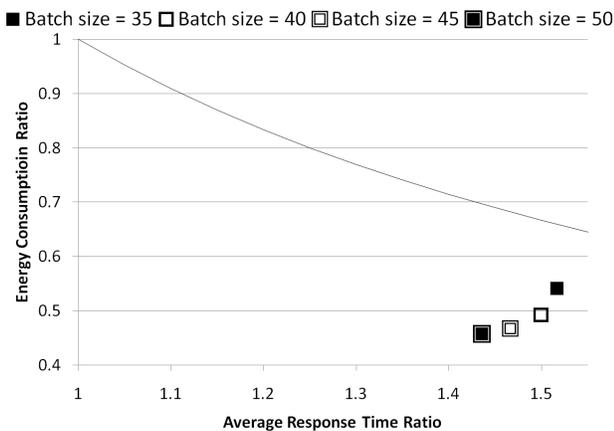

Figure 6: Energy versus Average per Query Response Time with QED

For future work, we plan on better understanding and measuring the energy consumption of all the hardware components in a DBMS server. This task is challenging as modern motherboards are complex (multi-layered) and tapping into the components (such as memory banks) is not trivial. Consequently, we may need to design some indirect ways of measuring the energy consumption of some components.

Designing a DBMS to balance the response time versus energy consumption opens a wide range of research issues that percolate through nearly all aspect of a DBMS, including query evaluation strategies, query optimization, query scheduling, physical database design, and dynamic workload management. In addition, there is also an opportunity for the database community to collaborate with the architecture community to influence the design of the next generation of hardware that will be more energy efficient, and to work towards building mechanisms that allow the DBMS to leverage the full potential of the energy saving features that the hardware will provide.

## 7. ACKNOWLEDGEMENT

We thank the anonymous CIDR reviewers for their insightful comments. We also thank Paul Beebe, Daniel Fabbri, Mark Hill, Steve Reinhardt, Thomas Wenisch, and Bruce Worthington for providing useful feedback on various parts of this work.

This research was supported by a grant from Microsoft. Any opinions, findings, and conclusions or recommendations expressed in this material are those of the authors and do not necessarily reflect the views of Microsoft.

## 8. REFERENCES

[1] Report To Congress on Server and Data Center Energy Efficiency. *EPA Technical Report*, 2007.
[2] L. A. Barroso and U. Hölzle. The Case for Energy-Proportional Computing. *IEEE Computer*, 40(12):33–37, 2007.
[3] C. Belady. In the Data Center, Power and Cooling Costs More than the IT Equipment it Supports. In *Electronics Cooling*, volume 23, 2007.
[4] K. G. Brill. Data Center Energy Efficiency and Productivity. In *The Uptime Institute - White Paper*, 2007.
[5] J. Dean and S. Ghemawat. MapReduce: Simplified Data Processing on Large Clusters. In *OSDI*, 2004.
[6] Enermax. http://www.enermaxusa.com/.
[7] X. Fan, W.-D. Weber, and L. A. Barroso. Power Provisioning for a Warehouse-Sized Computer. In *International Symposium on Computer Architecture*, pages 13–23, 2007.
[8] G. Graefe. Database Servers Tailored to Improve Energy Efficiency. In *Software Engineering for Tailor-made Data Management*, pages 24–28, 2008.
[9] S. Greenberg, E. Mills, W. Tschudi, P. Rumsey, and B. Myatt. Best practices for data centers: Lessons learned from benchmarking 22 data centers. *ACEEE Summer Study on Energy Efficiency in Buildings*, 2006.
[10] J. Hamilton. Where Does the Power go in DCs & How to get it Back. 2008.
[11] J. G. Koomey. Estimating Total Power Consumption by Servers in the US and the World. 2007.
[12] S. Rivoire, M. A. Shah, P. Ranganathan, and C. Kozyrakis. JouleSort: A Balanced Energy-Efficiency Benchmark. In *SIGMOD Conference*, pages 365–376, 2007.
[13] R. R. Schmidt, E. E. Cruz, and M. K. Iyengar. Challenges of Data Center Thermal Mangement. *IBM Journal of Research and Management*, 49(4), 2005.
[14] T. K. Sellis. Multiple-query optimization. *ACM Trans. Database Syst.*, 13(1):23–52, 1988.
[15] S. Siddha, V. Pallipadi, and A. V. D. Ven. Getting Maximum Mileage Out of Tickless. In *Proceedings of the Linux Symposium*, pages 201–208, June 2007.
[16] M. Stonebraker. Operating System Support for Database Management. *Communications of the ACM*, 24(7):412–418, 1981.
[17] Transaction Processing Council. http://www.tpc.org/tpch/.